\documentstyle[psfig]{mn}

\def\ga{\mathrel{\hbox{\rlap{\hbox{\lower4pt\hbox{$\sim$}}}\hbox{$>$}}}}

\title[SDSS J161033.64--010223.3: A second Cataclysmic Variable with a Non-radially Pulsating Primary]
{SDSS J161033.64--010223.3: A second Cataclysmic Variable with a Non-radially Pulsating Primary}
\author[Patrick A. Woudt and Brian Warner]
       {Patrick A. Woudt\thanks{E-mail: pwoudt@circinus.ast.uct.ac.za} 
        and Brian Warner\thanks{E-mail: warner@physci.uct.ac.za}\\
        Department of Astronomy, University of Cape Town, Private Bag,
        Rondebosch 7700, South Africa}
\date{18 June 2003}

\begin{document}

\maketitle

\begin{abstract}
We find from high speed photometry of the Sloan Digital Survey 
cataclysmic variable candidate SDSS J161033.64--010223.3 that it has a double 
humped modulation, which is probably orbital, with a period of 80.52 
minutes, and shows strong ZZ Cet type pulsations. The largest amplitude 
oscillations have periods near 607, 345, 304 and 221 s. The amplitudes of some of the 
oscillations are seen to vary on time scales of hours, indicating modes with 
unresolved multiplets in which the splittings are $\sim$ 1 d$^{-1}$. This is only the 
second dwarf nova found to have non-radial pulsations of its white dwarf 
primary.
\end{abstract}

\begin{keywords}
techniques: photometric -- binaries: close --
stars: individual:  SDSS J161033.64--010223.3, cataclysmic variables
\end{keywords}

\section{Introduction}

   The white dwarf accretors in cataclysmic variable (CV) stars have outer 
layers that differ from those of isolated white dwarfs. Not only is there an 
accumulation of hydrogen-rich material, leading ultimately to a nova 
explosion, the accretion process itself results in heating, mixing and 
compression which change the thermal structure and composition profile 
(Townley \& Bildsten 2003). Observations that contribute towards probing 
these structures are not easy to interpret -- the chemical compositions of the 
outer layers of the primaries of CVs are complicated by the rapid settling 
rate of the heavier elements (Sion 1999), the chemical abundances in nova 
ejecta are the end product of a mixture of dredge-up from the core and 
nuclear reactions during the explosion, embedded in shock-driven 
hydrodynamic processes. It would be of greater value to probe the outer 
layers by more direct means, as is possible through the application of 
asteroseismology to single white dwarfs (e.g., Clemens 1993). 

     A start in this direction was made by the discovery of the first CV to 
show non-radial pulsations (i.e. a CV whose primary is a ZZ Cet star) -- 
namely the dwarf nova GW Lib (Warner \& van Zyl 1998). GW Lib is a low 
mass transfer ($\dot{M}$) system that has only had one observed outburst (in 
1983) of very large amplitude. Its oscillation spectrum is complicated and 
has yet to be fully unravelled (van Zyl et al.~2000), and as such it remains to 
make a quantitative contribution to the determination of outer structure of an 
accreting white dwarf. Its predominant oscillation modes have periods that 
lie near 236, 376 and 650 s, with amplitudes in the range 5 -- 15 mmag, and 
evidence for other modes, including sum and difference frequencies, and 
some possible periods around 1000 s and 5000 s. As with large amplitude 
ZZ Cet stars, the oscillation spectra are not stationary, individual modes 
(which may in many cases be unresolved multiplets) varying greatly in 
amplitude on time scales of days, months and years.

    Nevertheless, the discovery of further examples of this class of CV will 
surely be welcomed, and that is what we present here -- a second CV in 
which large amplitude non-radial oscillations of the primary can be studied 
by the brightness modulations that they cause. In Section 2 we describe what 
was previously known about the star, in Section 3 we present our 
photometric observations and their Fourier transforms and in Section 4 we 
add some general comments.

\section{The star SDSS J161033.64--010223.3}

\begin{figure*}
\centerline{\hbox{\psfig{figure=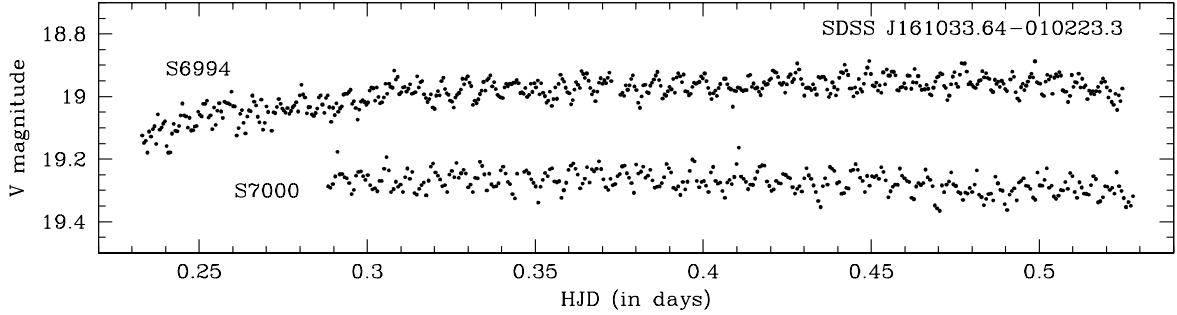,width=16.0cm}}}
  \caption{The light curves of SDSS1610 on 2003 June 7 and 9. Run S7000 has been displaced downwards by 0.3 mag for
display purposes only.}
 \label{lcsdss1610}
\end{figure*}

\begin{table*}
 \centering
  \caption{Observing log.}
  \begin{tabular}{@{}rrrrrcc@{}}
 Run No.  & Date of obs.          & HJD of first obs. & Length    & $t_{in}$ & Tel. &  V \\
          & (start of night)      &  (+2452000.0)     & (h)       &     (s)   &      & (mag) \\[10pt]
 S6998    & 2003 Jun 03 &  794.29592  &   1.92      &         60  &  74-in & 18.9  \\
 S6991    & 2003 Jun 06 &  797.26242  &   6.32      &         60  &  74-in & 18.9  \\
 S6994    & 2003 Jun 07 &  798.23303  &   7.00      &         45  &  74-in & 19.0  \\
 S6998    & 2003 Jun 08 &  799.47233  &   1.08      &         60  &  74-in & 18.9  \\
 S7000    & 2003 Jun 09 &  800.28837  &   5.75      &         60  &  74-in & 19.0  \\[5pt]
\end{tabular}
{\footnotesize 
\newline 
Note: $t_{in}$ is the integration time.\hfill}
\label{tab1}
\end{table*}

    The Sloan Digital Sky Survey has released an initial tranche of newly 
discovered CVs, found as candidates from their colours and then selected on 
the basis of spectroscopic appearance (Szkody et al.~2002). Among the 
multitude of spectra one in particular caught our attention. The nineteenth 
magnitude SDSS J161033.64-010223.3 (which we will abbreviate to SDSS1610) 
shows a blue continuum with broad hydrogen absorption lines containing 
narrow emission cores, and He\,I 5876 {\AA} weakly in emission. It is typical of 
a dwarf nova in quiescence. The depth of the H$\beta$ absorption, when compared 
with that for isolated white dwarfs (Wesemael et al.~1993), shows that there 
can be little overlying emission continuum from an accretion disc. The 
system is therefore in a very low state of mass transfer, a requirement if the 
heating effect of accretion is not to move the white dwarf out of the ZZ Cet 
instability strip (which for isolated DA white dwarfs is $\sim$ 11000 -- 12000 K 
(Bergeron et al.~1995)).
   
\section{Photometric observations}

   We have carried out high speed photometric observations of SDSS1610 
with the 74-inch Radcliffe reflector at the Sutherland site of the South 
African Astronomical Observatory, using the University of Cape Town CCD 
Photometer (O'Donoghue 1995) with no photometric filter (i.e., in `white 
light'). Our observing log is given in Table 1.

\begin{figure}
\centerline{\hbox{\psfig{figure=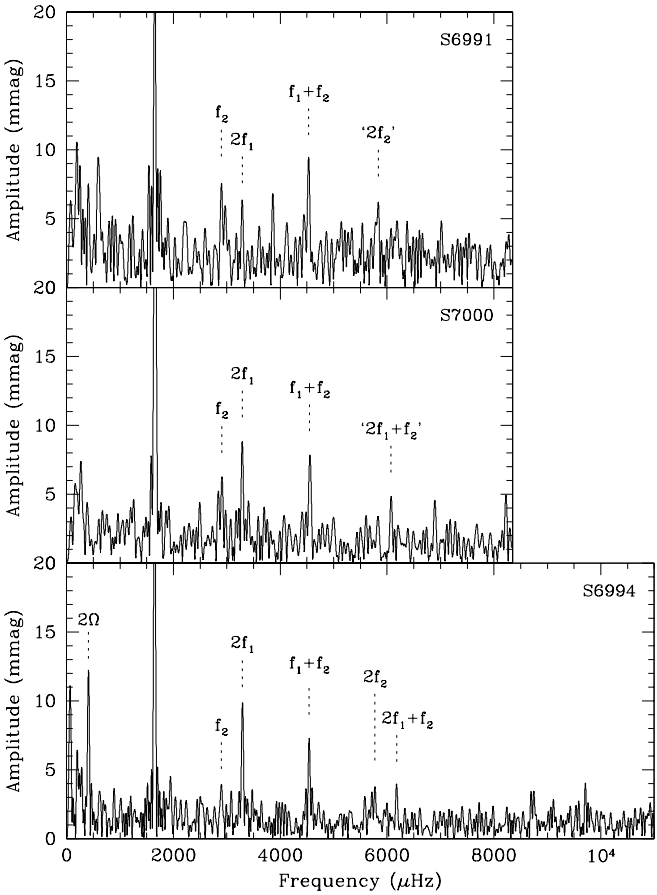,width=8.8cm}}}
  \caption{The Fourier transform of SDSS1610 for run S6991 (upper panel), run S7000 (middle panel)
and run S6994 (lower panel). The frequencies listed in Table 2 are marked here by label and dotted vertical
bars.}
 \label{ftsdss1610}
\end{figure}

    Variations of brightness on time scales $\sim$ 100s of seconds were evident 
even on the first night of observation, which was of poor photometric 
quality. In Fig.~\ref{lcsdss1610} we show the light curves obtained on later nights, of 
better quality. The slow variation is caused by differential extinction relative to the
(redder) comparison star. The presence of continuous oscillations with a range $\sim$ 0.10 
mag is obvious. Fourier transforms (FTs) of the three long runs (S6991, S6994 and S7000) 
with first and second order trends removed, are shown 
seperately in Fig.~\ref{ftsdss1610}. The dominant regions of oscillation power, in order of decreasing 
amplitude, are at $\sim$ 607 s, 2488 s (or its alias at 2419 s), 304 s (the first 
harmonic of 607 s), 220 s and 345 s, with amplitudes in the range 5 -- 25 
mmag. The frequencies are listed in Table 2. The fact that these periods bear no simple relationship to each other (as could be 
the case if the star was an intermediate polar in a low state) and their time scales
show that SDSS1610 is a second 
example of a CV with a non-radially oscillating primary. The absence of 
much overlying emission from the accretion disc implies that the observed 
amplitudes are not greatly reduced from those originating in the primary.

\begin{table*}
 \centering
  \caption{The suite of frequencies observed in SDSS J161033.64--010223.3.}
  \begin{tabular}{@{}rccrccrccrl@{}}
      &  \multicolumn{3}{c}{S6991}    & \multicolumn{3}{c}{S6994}    & \multicolumn{3}{c}{S7000}    & Ratio \\
 ID   & Frequency & Period           & Ampl.     & Frequency & Period        & Ampl.     & Frequency & Period    & Ampl. &  \\
      & $\mu$Hz   & s              & mmag    & $\mu$Hz   &  s          & mmag    & $\mu$Hz   & s       & mmag  &  \\[10pt]
$f_{1}$                 & 1648.9 $\pm$ 2.1 &   606.5 $\pm$ 0.8 & 23.4 & 1646.3 $\pm$ 1.2 & 607.4 $\pm$ 0.4 & 25.9 &  1651.0 $\pm$ 1.3 & 605.7 $\pm$ 0.5 & 30.0 & 1.000 \\
$2\,f_{1}$             & 3286.4 $\pm$ 9.1 &   304.3 $\pm$ 0.8 &  6.3 & 3292.3 $\pm$ 2.9 & 303.7 $\pm$ 0.3 & 10.0 &  3290.6 $\pm$ 3.9 & 303.9 $\pm$ 0.4 &  9.7  & 1.995 \\
$f_{2}$                 & 2899.0 $\pm$ 7.7 &   344.9 $\pm$ 0.9 &  7.5 & 2896.0 $\pm$ 7.5 & 345.3 $\pm$ 0.9 &  4.0 &  2907.3 $\pm$ 6.8 & 344.0 $\pm$ 0.8 &  6.3 & 1.759 \\
$f_{1} + f_{2}$         & 4530.7 $\pm$ 6.1 &   220.7 $\pm$ 0.3 &  9.4 & 4540.0 $\pm$ 4.1 & 220.3 $\pm$ 0.2 &  7.2 &  4547.3 $\pm$ 4.6 & 219.9 $\pm$ 0.2 &  8.2 & 2.753 \\
$2\,f_1 + f_2$         &                  &                   &      & 6177.6 $\pm$ 7.6 & 161.9 $\pm$ 0.2 &  4.0 &  6075.2 $\pm$ 8.6:& 164.6 $\pm$ 0.2:&  5.0: & 3.747 \\
$2\,f_{2}$             & 5834.7 $\pm$ 9.3:&   171.4 $\pm$ 0.3:&  6.2:& 5771.6 $\pm$ 8.0 & 173.3 $\pm$ 0.2 &  3.7 &                   &                 &      &\\
$\Omega$          &\hfill 190.5 $\pm$ 4.0 &  5249 $\pm$ 110   & 11.0 &\hfill  198.7 $\pm$ 11. & 5033 $\pm$ 279  & 10.1 &                   &                 &  &    \\
$2\,\Omega$      &\hfill 406.9 $\pm$ 6.1 &   2458 $\pm$ 37   &  7.5 &\hfill  410.2 $\pm$ 2.2 & 2437 $\pm$ 13   & 11.9 &                   &                 &    &  \\[5pt]

\end{tabular}
{\footnotesize 
\newline 
Note: `:' indicates an uncertain identification.\hfill}
\label{tab2}
\end{table*}

\begin{figure}
\centerline{\hbox{\psfig{figure=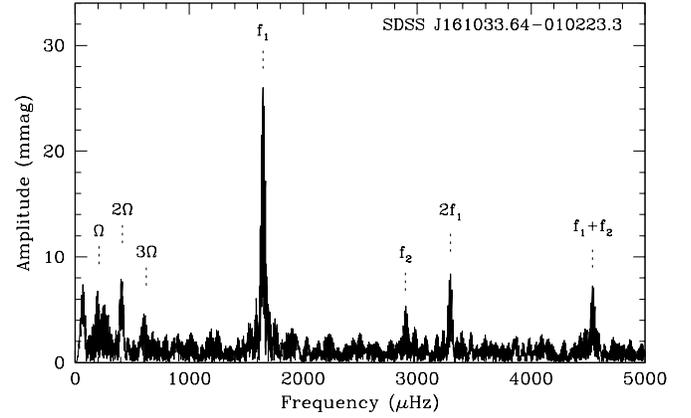,width=8.8cm}}}
  \caption{The Fourier transform of the three long runs combined, showing the low frequency range.}
 \label{ftall}
\end{figure}

\begin{figure}
\centerline{\hbox{\psfig{figure=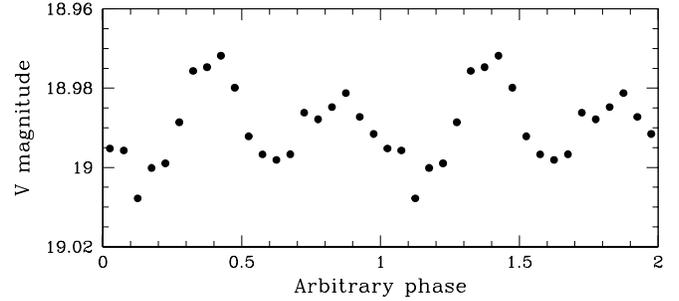,width=8.8cm}}}
  \caption{The average light curve of SDSS1610 (run S6994) folded on the 80.52 min period, after the light curve
had been prewhitened at $f_{1}$, $2\,f_{1}$ and $f_{1} + f_{2}$.}
 \label{lcorb}
\end{figure}

    There are low frequency peaks in the FTs of individual nights in the 
vicinity of 200 $\mu$Hz and its first and second harmonics. In the FT of the three 
long runs combined (Fig.~\ref{ftall}) there is a choice of aliases, the best-fitting 
being 207.0 ($\pm$ 0.4), 413.7 ($\pm$ 0.2) and 625.1 ($\pm$ 0.6) $\mu$Hz. This is equivalent 
to a period of 80.52 min and its harmonics. The alternate choice is 
equivalent to 85.08 min. We suggest that these features are produced by 
orbital modulation. As has been pointed out by Thorstensen et al.~(2002), the 
dwarf novae with orbital periods in the vicinity of 80 mins are a particularly 
interesting group, of which GW Lib ($P_{orb}$ = 76.78 min from spectroscopic 
observations) is a member. The orbital signal is strongest in run S6994. We 
have prewhitened the light curve at the principal oscillation frequencies and 
produced a light curve folded on the 80.52 period, which is shown in Fig.~\ref{lcorb}.
It is a double humped light curve, very similar to that of WZ Sge 
(Patterson 1980) and WX Cet (Rogoziecki \& Schwarzenberg-Czerny 2001), 
both of which are very low $\dot{M}$ dwarf novae where it is proposed that the 
reason for two humps per orbit is because the bright spot on the disc is 
visible from behind through the low optical thickness disc (Robinson, Nather 
\& Patterson 1978a).

The FTs of SDSS1610 bear a strong resemblence to that of the ZZ Cet star VY Hor (O'Donoghue, Warner
\& Cropper 1992), where the principal frequency is 1626 $\mu$Hz at an amplitude of 60 mmag. In VY Hor and
GD 154 (Robinson et al.~1978b), the eigenfrequencies are completely described by the sequences $pf$ and 
$(q+{1\over{2}}+\epsilon)f$, where $p$ = 1, 2, ..... and $q$ = 0, 1, 2, .....
For VY Hor $\epsilon = 0.037$ and for GD~154 $\epsilon = -0.03$. A similar situation obtains in PG 1351+489, where
$\epsilon = -0.03$ (Winget, Nather \& Hill 1987).

In the case of SDSS1610, where the amount of observational material is much less than was available for these
ZZ Cet star studies, we can fit the few observed eigenfrequencies with $\epsilon \approx 0.25$. In Table 2 the final
column gives the ratio of the average with respect to the mean $f_1$ of the frequencies measured on the three nights
(for the mean of `$2\,f_1+f_2$' we use only run S6994, in run S7000 the identification of this mode is uncertain).
It should be remembered that the peak frequencies quoted in Table 2 are those of unresolved multiplets and therefore
would be expected to vary slightly from one run to the next. For this reason, we have not given frequencies determined
from the FT of the combined runs, which in addition introduces the problem of aliasing.
If the proximity to values of $(q+{1\over{2}})f$ in GD 154, VY Hor and PG 1351+489 is a sign of non-linear
resonant coupling between modes (O'Donoghue et al.~1992), it would appear that in SDSS1610 we are seeing resonances
close to another commensurability, viz $(q+{3\over{4}})f$. In GW Lib, a similar structure is seen in the FT with the
presence of $q = 0, 1$ and 2 peaks, and $p = 1$ peaks, where $f = 1545$ $\mu$Hz and $(q+{3\over{4}}+\epsilon)f$ with
$\epsilon \approx -0.03$.

\begin{figure}
\centerline{\hbox{\psfig{figure=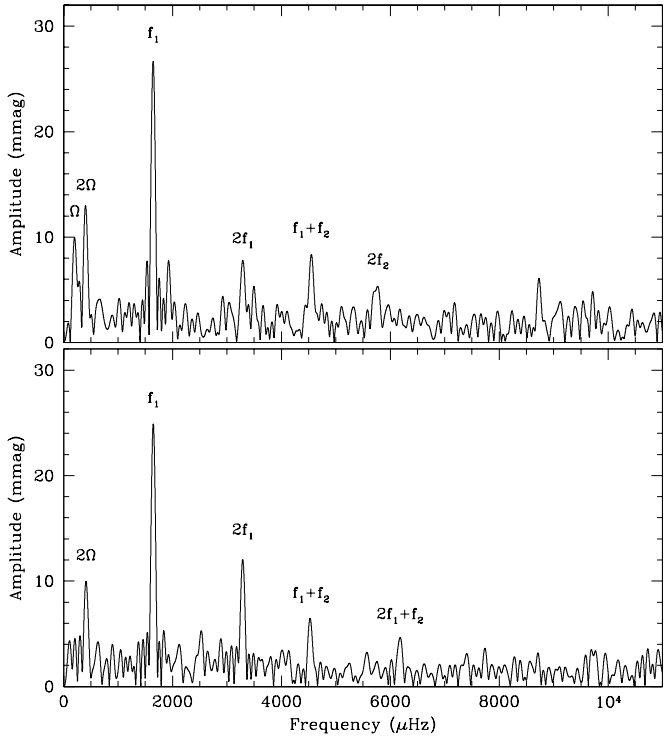,width=8.8cm}}}
  \caption{The Fourier transform of the first 3.5 h of run S6994 (upper panel) and the last 3.5 h of run S6994 (lower panel).
$\Omega$ is the orbital frequency (1/$P_{orb}$ = 207 $\mu$Hz).}
 \label{sdss1610ftbu}
\end{figure}

The amplitudes of the various eigenmodes vary from night to night and even within a run. In Fig.~\ref{sdss1610ftbu}
we show FTs of the first and second halves of run S6994 which illustrate that the first harmonic of the principal
eigenfrequency is varying in strength. As with other ZZ Cet stars, this simply means that there is unresolved
multiplet structure in these FTs.

A few unidentified low amplitude peaks ($\sim 4 - 5$ mmag) present in the FTs are not listed in Table 2; they occur in individual runs only.
These are: 3860.1 $\mu$Hz in run S6991; 8731.1 $\mu$Hz and 9712.3 $\mu$Hz in run S6994; 6893.6 $\mu$Hz and 8225.4 $\mu$Hz
in run S7000.

\section{discussion}

Clemens (1993) partitioned ZZ Cetis into a `hot' group and a `cool' group and 
shows that the distribution and range of periods is different in the two 
groups. 
A decade later, with far more ZZ Cet stars known, this division is less convincing; there are
many stars that combine oscillation modes from both groups, as do GW Lib and SDSS1610
Clemens also pointed out 
that similarities in the ZZ Cetis, at least among the hotter ones, probably arise because 
all have a similar amount of hydrogen, $\sim 10^{-4}$ M$_{\odot}$ in their envelopes. The 
primary of a CV could have any amount of hydrogen, from zero to $\sim 10^{-4}$ 
M$_{\odot}$, depending on when it last had a nova eruption.

SDSS1610 probably has rare dwarf nova outbursts of amplitude similar to WZ Sge and would therefore
reach $m_V \sim 11$ at maximum. A search through archived sky patrol plates is therefore recommended,
and a watch should be kept for future outbursts. The value of following the heating effects of an outburst
on the eigenfrequencies of the primary has already been pointed out (Warner \& van Zyl 1998).

\section*{Acknowledgements}
PAW is supported by funds made available from the National Research
Foundation and by strategic funds made available to BW from the
University of Cape Town. BW's research is supported by the University.

\end{document}